\documentclass[a4paper,prb,aps,epsfig,twocolumn,showpacs,preprintnumbers,superscriptaddress,float,amsmath,amssymb]{revtex4}

\usepackage{graphicx}% Include figure files
\usepackage{dcolumn}% Align table columns on decimal point
\usepackage{bm}% bold math
\usepackage{graphicx}
\usepackage{pgfplots}
\usetikzlibrary{arrows.meta}
\pgfplotsset{compat=1.16}

\newcommand{\sgn}{{\rm sgn}}

\begin{document}

\title{Topology shared between classical metamaterials and interacting superconductors}
%\title{Metameterials that behave like strongly coupled metals and superconductors}
%\title{Mapping strongly coupled superconductors to nonlinear topological mechanics}
\date{\today}

\author{Po-Wei Lo}
\affiliation{Laboratory of Atomic and Solid State Physics,
Cornell University, Ithaca, NY, 14853}
\author{Chao-Ming Jian}
\affiliation{Laboratory of Atomic and Solid State Physics,
Cornell University, Ithaca, NY, 14853}
\author{Michael J Lawler}
\affiliation{Laboratory of Atomic and Solid State Physics,
Cornell University, Ithaca, NY, 14853}
\affiliation{Department of Physics, Applied Physics, and Astronomy, Binghamton University, Binghamton, New York 13902}

\begin{abstract}

Supersymmetry
%Many-body supersymmetric structure 
has been studied at a linear level between normal modes of metamaterials described by 
%through an analogy between 
rigidity matrices and non-interacting quantum Hamiltonians. The connection between classical and quantum was made through the matrices involved in each problem. Recently, %advancement has been made to 
insight into the behavior of nonlinear mechanical systems was found by defining topological indices via the Poincar\'e-Hopf index. It turns out, because of the mathematical similarity, 
%in mathematical frameworks, 
this topological index shows a way to approach supersymmetric \emph{quantum theory} from classical mechanics. 
Using this mathematical similarity, we establish a topological connection between
% , we study two topological numbers in 
isostatic mechanical metamaterials and supersymmetric quantum systems, such as electrons coupled to phonons in metals and superconductors. Firstly, we define $Q_{net}$ for an isostatic mechanical system that counts the minimum number of zero-energy configurations. Secondly, we write a supersymmetric Hamiltonian that describes a metal or a superconductor interacting with anharmonic phonons.
%a metal or superconductor strongly coupled to anharmonic phonons %\cmj{[Comment: maybe change ``a metal or superconductor strongly coupled to anharmonic phonons" to just "a superconductor interacting with anharmonic phonons"]} . 
This Hamiltonian has a Witten index, a topological invariant that captures the balance of bosonic and fermionic zero-energy states.  
We are able to connect these two systems by showing that $Q_{net}=W$ under very general conditions. 
Our result shows that (1) classical metamaterials can be used to study the topology of 
%strongly coupled 
interacting quantum systems with aid of supersymmetry, and
(2) with fine-tuning between anharmonicity of phonons and couplings among Majorana fermions and phonons, it is possible to realize such a supersymmetric quantum system that shares the same topology as classical mechanical systems.
%, and (3) there is an alternative way to understand the topology of supersymmetric quantum systems from the angle of classical mechanics. 
%\cmj{[Comment: I think (3) is identical to (2)]}

\end{abstract}

\maketitle

\section{Introduction}

Mechanical systems offer concrete models to understand abstract ideas in physics. Recently, through the analogy between rigidity matrices and non-interacting quadratic Hamiltonians\cite{PhysRevB.94.165101,PhysRevResearch.1.032047,PhysRevResearch.3.023213}, advancement has been made in the fields of topological metamaterials\cite{lawler2013emergent,tm1,tm2,tm3,tm4,tm5} and topological magnets\cite{origami2,roychowdhury2018classification}. Beyond the linear level, a prescription of defining topological indices via the Poincar\'e-Hopf index has been introduced to study topology in nonlinear mechanical systems\cite{PhysRevLett.127.076802}. This prescription gives a hint that there exist some topological connections between nonlinear mechanical systems and supersymmetric quantum systems due to their similar mathematical frameworks\cite{PhysRevLett.127.076802,10.4310/jdg/1214437492,tft}.

The topological index $\mu(p)$ defined for a zero-energy configuration point $p$ in an isostatic mechanical system has a similar mathematical expression that is used to calculate the supersymmetric partition function in the topological quantum field theory\cite{PhysRevLett.127.076802,tft}. For a certain ``symmetric" case, the sum over all $\mu(p)$ is exactly equal to the Witten index of the BRST type supersymmetric model\cite{brst1,brst2,brst3}. The definition of this ``symmetric" case will be provided below. In this model, the Hamiltonian can be interpreted as complex fermions that conserve fermion numbers, such as electrons in a normal metal, coupled to anharmonic phonons. 
However, in a mechanical system, constraints, in general, do not have this symmetry. So this connection seems restricted to limited cases.

Fortunately, a more general supersymmetric Hamiltonian can be written in a way that does not require the constraint functions to obey this symmetry\cite{PhysRevB.94.165101}. In this case, $U(1)$ symmetry of the fermion systems is broken and thus the Hamiltonian can then be interpreted as Majorana fermions (which can realize a $p$-wave superconductor\cite{Kitaev_2001}) coupled to anharmonic phonons. Although the fermion number is no longer conserved, the fermion parity is still well-defined. Thus, we can calculate the Witten index even in a ``non-symmetric'' case. Then a question arises: for a generic set of constraint functions, what is the relation between the Witten index and the topological index $\mu(p)$?

\begin{figure}
  \includegraphics[width=8cm]{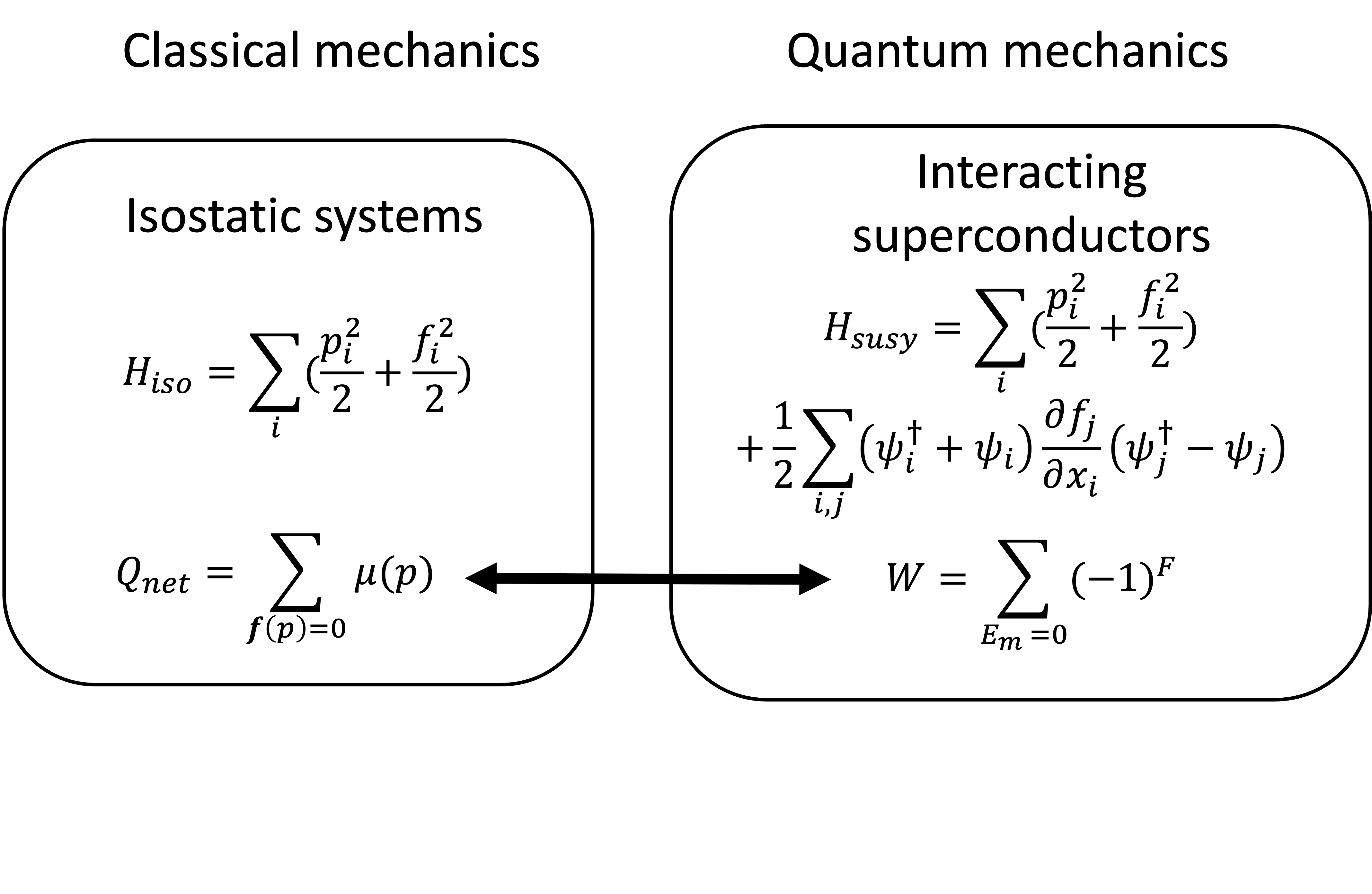}
  \caption{Topology shared between classical metamaterials and interacting superconductors via the topological index $Q_{net}$ and the Witten index $W$.}
  \label{fig:map}
\end{figure}

To answer this question, we study the topology shared between classical constraint problems and interacting metals or superconductors. Firstly, we define $Q_{net}$ for an isostatic mechanical system as the sum over all $\mu(p)$ and find that its magnitude is the minimum number of zero-energy configurations. Secondly, we write a supersymmetric Hamiltonian that has a well-defined Witten index $W$ for a generic set of nonlinear constraint functions. We show that this Hamiltonian can describe a superconductor interacting with phonons, including any anharomonicity they may have. $|W|$ for this Hamiltonian also turns out to be the minimum number of zero-energy states. Finally, we make a topological connection between these two systems by showing that $Q_{net}=W$ for a set of nonlinear and non-symmetric constraints under very general conditions (specified below) as shown in Fig.\ref{fig:map}.

\section{zero-energy configurations in an isostatic mechanical system}

Firstly, we consider an isostatic mechanical system described by a Hamiltonian
\begin{equation}
    H_{iso} = \sum_i \left(\frac{p_i^2}{2}+\frac{f_i^2}{2}\right)
\label{eq:Hiso}    
\end{equation}
which has zero-energy configurations satisfying a set of constraints $f_1=0,f_2=0,...,f_n=0$ where $f_i$ is a function of $x_1,x_2,...,x_n$ such as those that arise in {\it e.g.} springs, linkages, and origami. When $f_i$ is a linear function, $H_{iso}$ describes $n$ simple harmonic oscillators. Following the definition in Ref.\cite{PhysRevLett.127.076802}, a topological index $\mu(p)$ at a zero-energy configuration $p$ can be calculated by an integration of a differential form
\begin{equation}
    \label{eq:windingnumber}
    \mu(p) = \frac{1}{s_{n-1}(n-1)!}\oint_{S_p} \frac{f_{i_1} d f_{i_2}\wedge ... \wedge d f_{i_n} \epsilon^{i_1,i_2,...,i_n}}{(f_1^2+f_2^2+...+f_n^2)^{\frac{n}{2}}}
\end{equation}
where $S_p$ is an $(n-1)$-dimensional sphere in the configuration space which encloses the point $p$, $s_{n-1}$ is the surface area of a unit $(n-1)$-dimensional sphere. When the Jacobian $\partial f_{i} / \partial x_j$ at $p$ is full rank, $\mu(p) = \textrm{sgn}[\textrm{det}( \partial f_{i} / \partial x_j )]$.

Here we further define another topological index $Q_{net}$ as the sum over $\mu(p)$ of all zero-energy configurations.
\begin{equation}
    Q_{net} = \sum_{\mathbf{f}(p)=\mathbf{0}}\mu(p)=\sum_{\mathbf{f}(p)=\mathbf{0}}\textrm{sgn}\left[\textrm{det}\left(\frac{ \partial f_{i} }{ \partial x_j} \right)\Bigg|_p \right]
\label{eq:Qnet}
\end{equation}
which counts the difference between the number of zero-energy configurations with $\mu=+1$ and $\mu=-1$. Because $\mu$ can only be created or annihilated in pairs, $|Q_{net}|$ is the minimum number of zero-energy configurations that always exist under finite local deformations.

\begin{figure}
  \includegraphics[width=8cm]{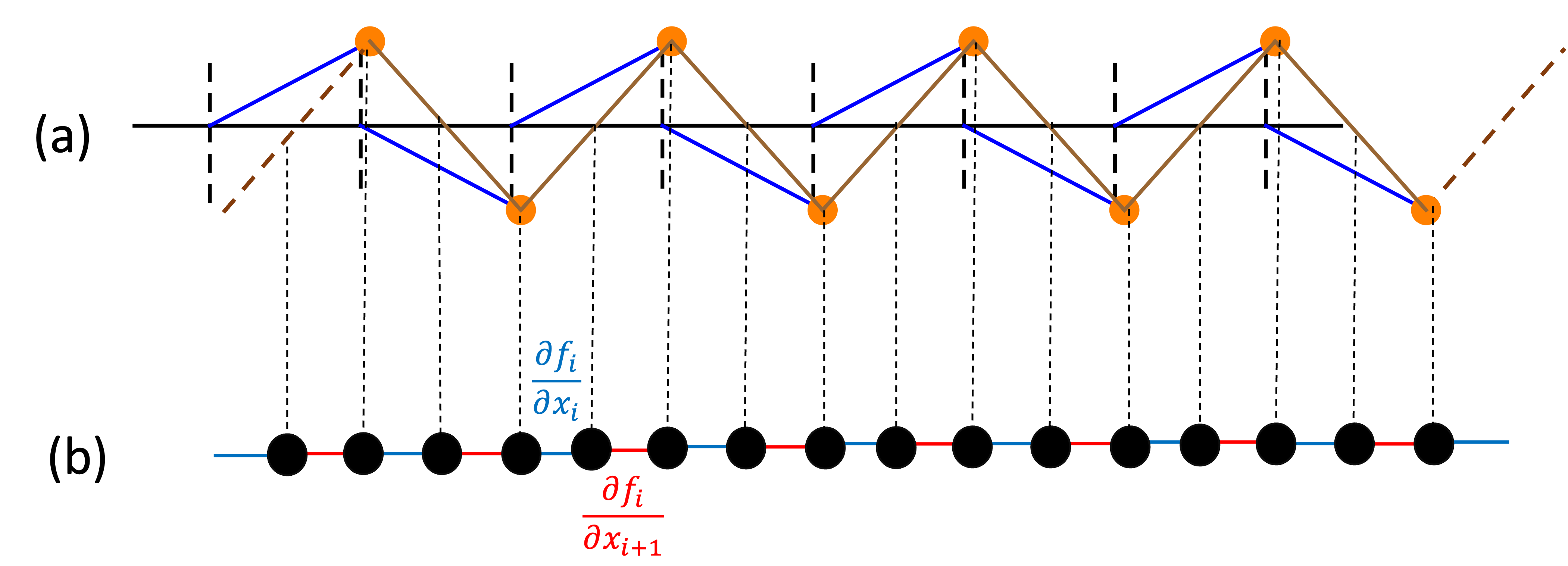}
  \caption{(a)The Kane-Lubensky chain with periodic boundary conditions. (b)The Kitaev chain. For each constraint (spring) or degree of freedom (ball) of the KL chain, we put a Majorana fermion that hops to its nearest neighbors with the parameters $\frac{\partial f_i}{\partial x_i}$ and $\frac{\partial f_i}{\partial x_{i+1}}$.}
  \label{fig:eg}
\end{figure}

%Let's consider some examples in mechanical systems. A one-dimensional chain contains $n$ balls and $n$ spring sequentially is depicted in Fig.\ref{fig:eg}(a). In this case, all constraints are linear, and thus $Q_{net}=\pm 1$. This implies that there always exist one solution for any choices of spring lengths.

Let's consider an example, the Kane-Lubensky(KL) chain\cite{tm1} with periodic boundary conditions as shown in Fig.\ref{fig:eg}(a). This example is an anharmonic-oscillator system that naturally exists in nonlinear mechanical systems. There are many zero-energy configurations, but the sum over $\mu(p)$ is zero. Therefore, $Q_{net}=0$ suggests that all zero-energy configurations can be annihilated by deforming constraints. For example, if we choose one of the spring lengths larger than twice the length of rotors plus the distance between the two nearest pivot points, then there will be no zero-energy configuration in the KL chain.

\section{zero-energy states in a supersymmetric quantum system}

Secondly, we consider a supersymmetric quantum system similar to Ref.\cite{PhysRevB.94.165101} described by a supersymmetric Hamiltonian
\begin{equation}
    H_{susy} = \{Q,Q\}
\end{equation}
where $Q=\frac{1}{2}\sum_{i}[\psi_i(p_i+if_i)+\psi_i^{\dagger}(p_i-if_i)]$ and $\psi_i$ is a fermion operator. In the Euclidean quantum theory, we can replace $p_i$ by $i\frac{\partial}{\partial x_i}$. Then $H_{susy}$ can be rewritten as
\begin{equation}
    H_{susy} = \sum_i \left(\frac{p_i^2}{2}+\frac{f_i^2}{2} \right) + \frac{1}{2}\sum_{i,j}(\psi_i^{\dagger}+\psi_i)\frac{\partial f_j}{\partial x_i}(\psi_j^{\dagger}-\psi_j)
    \label{eq:Hsusy}
\end{equation}
which can also be written in terms of Majorana fermion operators $\gamma_{a,i}=\psi_i^{\dagger}+\psi_i$ and $\gamma_{b,i}=-i(\psi_i^{\dagger}-\psi_i)$ as
\begin{equation}
    H_{susy} = \sum_i\left(\frac{p_i^2}{2}+\frac{f_i^2}{2}\right) + \frac{i}{2}\sum_{i,j}\gamma_{a,i}\frac{\partial f_j}{\partial x_i}\gamma_{b,j}
\label{eq:Hsusy}    
\end{equation}
Here we can see that $H_{susy}$ and $H_{iso}$ only differ by additional terms described by the interacting between fermions and bosons. When $f_i$ is a linear function, $H_{susy}$ is simply two independent systems, $n$ simple harmonic oscillators and a non-interacting Majorana fermion system. In general, a constraint function $f_i$ is nonlinear. We can get some insights by expanding $f_i$ around a zero-energy configuration point to second highest order terms ($f_i = \sum_{j}a_{i,j}x_j + \sum_{j,k} b_{i,j,k}x_j x_k$). By doing so, we will get
\begin{equation}
\begin{split}
    \tilde{H}_{susy} = \sum_i \left(\frac{p_i^2}{2}+\frac{(\sum_{j} a_{i,j}x_j+\sum_{j,k}b_{j,i,k}x_j x_k)^2}{2}\right)\\
    + \frac{i}{2}\sum_{i,j}\gamma_{a,i}a_{j,i}\gamma_{b,j}
    + \frac{i}{2}\sum_{i,j,k}\gamma_{a,i}b_{j,i,k}x_k\gamma_{b,j}
\end{split}
\end{equation}
The first and second terms describe anharmonic phonons and a non-interacting Majorana fermion system, respectively, and the last term is the coupling between Majorana fermions and anharmonic phonons.

In the supersymmetric quantum system, nonzero-energy states are always paired with opposite fermion parities. Thus, we can calculate the Witten index 
\begin{equation}
    W = \sum_{E_{m}=0}(-1)^{F}
\label{eq:Witten}
\end{equation}
where $E_m$ is an eigenenergy of $H_{susy}$ and $(-1)^F$ is the fermion parity operator. Because the Witten index tells us the difference between the number of even and odd fermion parity zero-energy states, its magnitude $|W|$ is the minimum number of zero-energy states that always exist under finite local deformations. 

In a symmetric case where $\partial f_i/\partial x_j$ is a symmetric matrix (with respect to the matrix indices $i,j$). We can find a function $V$ such that $f_i = \frac{\partial V}{\partial x_i}$. $H_{susy}$ is reduced to
\begin{equation}
    H_{susy}^{sym} = \sum_i \left(\frac{p_i^2}{2}+\frac{f_i^2}{2}\right) + \frac{1}{2}\sum_{i,j}\frac{\partial f_j}{\partial x_i}(\psi_i\psi_j^{\dagger}-\psi_i^{\dagger}\psi_j)
\end{equation}
whose path integral can be viewed as a Witten-type supersymmetric topological quantum field theory\cite{tft}. Similar to Eq.\ref{eq:Hsusy}, but now it describes fermions (electrons in a metal) coupling to anharmonic phonons. In this case, it has been shown that $W=\sum_{\frac{\partial V}{\partial x_i}=0} {\rm sgn}[\det(\frac{\partial^2 V}{\partial x_i \partial x_j}\Big|_p)]$ which is exactly the same as $Q_{net}$.

Given those similar physical interpretations of $Q_{net}$ and $W$ plus the result of symmetric cases, it seems that $Q_{net}$ might still be related to $W$ in a certain way even for non-symmetric cases.

\subsection{linear functions}

\begin{figure}
  \includegraphics[width=8cm]{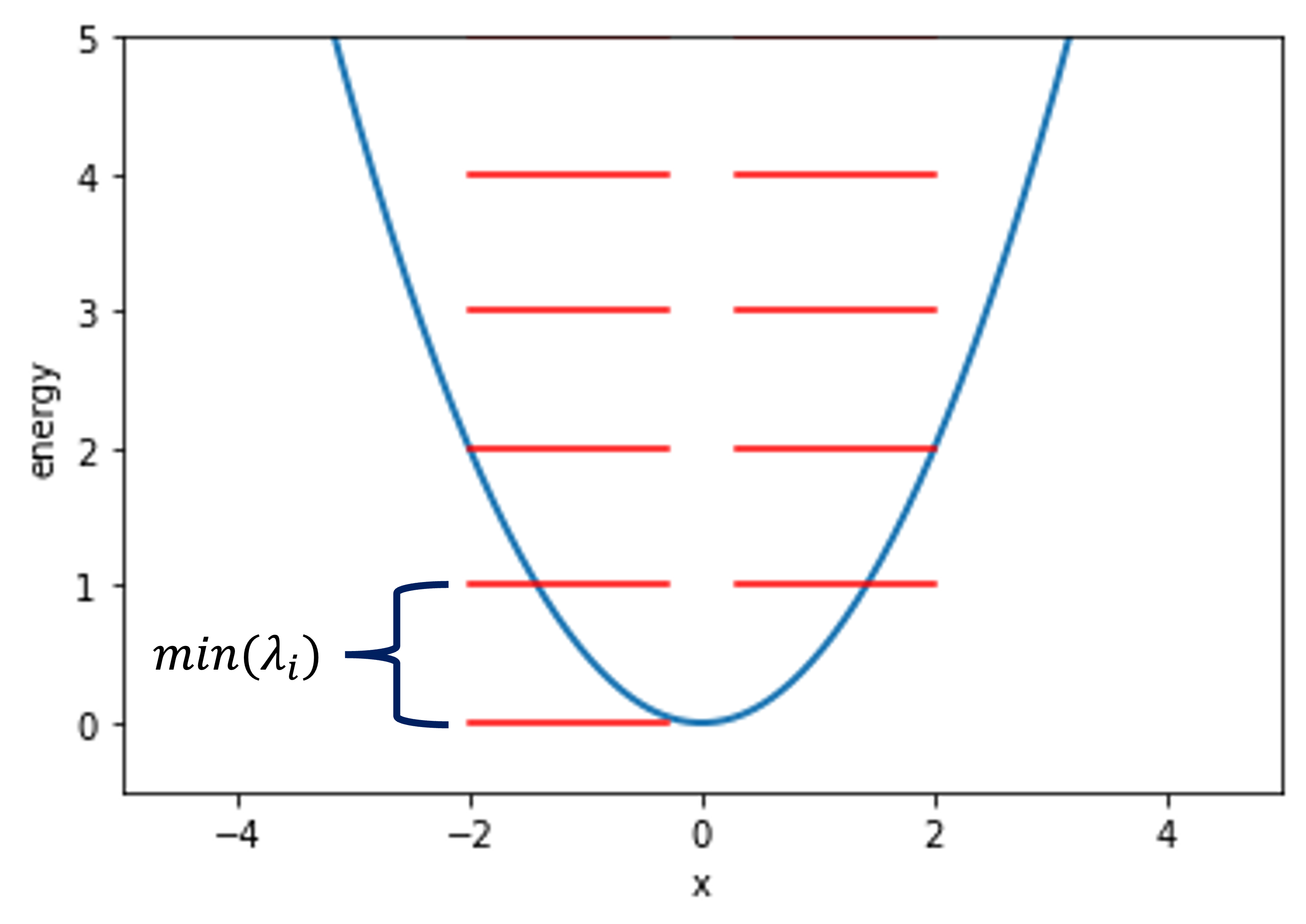}
  \caption{Spectrum of $H_{susy}^L$ with a single linear constraint function $f(x)=x$.}
  \label{fig:sp1}
\end{figure}

To find their connections, we first look at linear-constraint cases to get some insights. Assume that the constraints are $\mathbf{Rx=0}$. Then the corresponding $H_{susy}$ is
\begin{equation}
    H_{susy}^{L} = \sum_i\frac{p_i^2}{2}+\sum_{i,j,k}\frac{x_i\mathbf{R}^{T}_{i,j}\mathbf{R}_{j,k}x_k}{2} + \frac{i}{2}\sum_{i,j}\gamma_{a,i}\mathbf{R}_{i,j}\gamma_{b,j}
\end{equation}
By performing the singular value decomposition to obtain $\mathbf{R=U\Sigma V^{T}}$, and rotating $x'_i=\mathbf{V}_{i,j}x_j$, $\gamma'_{a,i}=\mathbf{U}^{T}_{i,j}\gamma_{a,j}$ and $\gamma'_{b,i}=\mathbf{V}_{i,j}\gamma_{b,j}$, we get
\begin{equation}
    H_{susy}^{L} = \sum_i \left[\frac{(p'_i)^2}{2}+\frac{(\lambda_i x'_i)^2}{2} + \frac{i\lambda_i}{2} \gamma'_{a,i}\gamma'_{b,j}\right]
\label{eq:HL}
\end{equation}
where $\lambda_i=\mathbf{\Sigma}_{i,i}$ is the singular value of $\mathbf{R}$.
$H_{susy}^L$ contains two non-interacting systems. The first one described by the first two terms in Eq.\ref{eq:HL} is $n$-simple-harmonic-oscillators with ground state energy equal to $\sum_i \frac{\lambda_i}{2}$. The last term is a Majorana fermion system that has energy $\sum_i \pm \frac{\lambda_i}{2}$ in different Majorana fermion sectors. The ground state energy of the Majorana fermion system is $-\sum_i \frac{\lambda_i}{2}$ and its fermion parity equates to the Pfaffian of the Hamiltonian in the Majorana basis which is simply $\sgn[\det(\mathbf{R})]$\cite{Kitaev_2001}. Combining the two systems, the lowest energy state of $H_{susy}^L$ has exact zero energy with a gap $\min(\{\lambda_i\})$ as shown in Fig.\ref{fig:sp1}. Because $H_{susy}^L$ only has one single zero-energy state, $W=\sgn[\det(\mathbf{R})]$ which is equal to $Q_{net}$.  

For example, in the periodic KL chain, if we linearize the constraints at a uniform solution point\cite{tm1,PhysRevLett.127.076802}, $\frac{\partial f_i}{\partial x_i}$ and $\frac{\partial f_i}{\partial x_{i+1}}$ would be constants. In the fermionic part, we will get the Kitaev chain as shown in Fig.\ref{fig:eg}(b) which is a $p$-wave superconductor\cite{Kitaev_2001}. 

\subsection{nonlinear functions}

Now let's go back to generic nonlinear-constraint cases. Here we specify three general conditions for the constraint functions $\mathbf{f}$ that we are interested in. (1)$\frac{\partial f_i}{\partial x_j}$ is continuous everywhere. This makes sure that potential energy is continuous in the whole space.(2)$||\mathbf{f}|| \rightarrow \infty$ as $||\mathbf{x}|| \rightarrow \infty$. This guarantees that wavefunctions are confined in finite regions. (3)The Jacobian $\frac{\partial f_i}{\partial x_j}$ is full rank at all solution points $\mathbf{f=0}$. 

To find $W$, we rescale the constraint functions $f_i$ by a positive constant $g$ and rewrite the Hamiltonian as
\begin{equation}
    H_{susy}(g) = \sum_i \left(\frac{p_i^2}{2}+\frac{g^2f_i^2}{2}\right) + \frac{ig}{2}\sum_{i,j}\gamma_{a,i}\frac{\partial f_j}{\partial x_i}\gamma_{b,j}
\end{equation}
and first look at large $g$ cases.

%We can do similar rescaling on the classical side and get $H_{iso}(g)=\sum_i(\frac{p_i^2}{2}+\frac{g^2f_i^2}{2})$. Classically, the large $g$ limit corresponds to the constraints made by stiff springs or mechanical linkages.
When $g$ is very large, the potential energy is dominated by $\frac{g^2f^2_i}{2}$ term. Thus, we can focus on those points where $\mathbf{f=0}$ to study low-energy states. Assume that we have $N$ points satisfying $\mathbf{f=0}$ labeled as $z_{\alpha=1,2,...,N}$. At each $z_\alpha$, we take the linear order of $f_i$ to obtain a Hamiltonian which locally looks like a potential well described by Eq.\ref{eq:HL}. The structure of the low energy states is, therefore, similar to a linear-constraint case in which a system has a zero-energy state gapped by $\min(\{g\lambda_{\alpha,i}\})$ where $\lambda_{\alpha,i}$ is a singular value of the matrix $\frac{\partial f_i}{\partial x_j}$ at point $z_\alpha$.

Two types of perturbations can lift or lower energy. Firstly, we consider the overlap between two wave functions localized at different $z_\alpha$. The overlap is estimated as $\sim e^{-\epsilon g}$ where $\epsilon$ is some positive constant that depends on the distance between two wells. Thus, the energy will only be increased or lowered by an amount of order of $~e^{-\epsilon g}$. 

\begin{figure}
  \includegraphics[width=8cm]{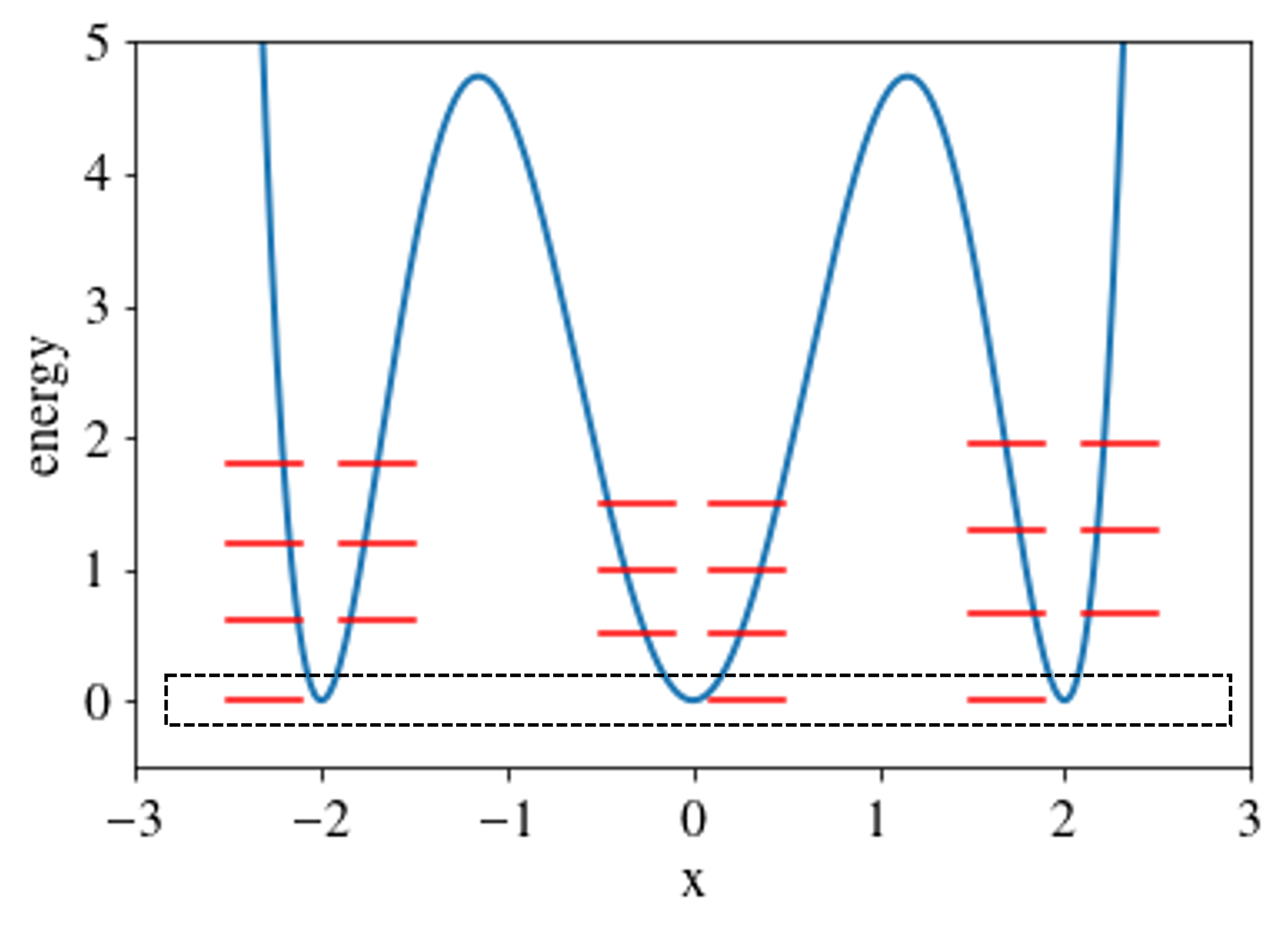}
  \caption{Low-energy spectrum of $H_{susy}(g)$ with nonlinear constraint functions in the large $g$ limit.}
  \label{fig:sp2}
\end{figure}

The second perturbation is the higher order corrections terms around each well. We expand $f_i$ at each $z_\alpha$ as $f_i = \sum_{j}a_{i,j,\alpha}x_j + \sum_{j,k}b_{i,j,k,\alpha}x_j x_k+...$ Then we rescale $x_i$ with a prefactor $g^{-1/2}$, namely, $x_i\rightarrow g^{-1/2} x_i$ and $p_i\rightarrow g^{1/2} p_i$. We get $gf_i \rightarrow g^{1/2}\sum_{j}a_{i,j,\alpha}x_j + \sum_{j,k}b_{i,j,k,\alpha}x_j x_k+...$, and $g\frac{\partial f_j}{\partial x_i} \rightarrow g a_{j,i,\alpha} + g^{1/2} \sum_{j}b_{j,i,k,\alpha} x_k+...$. As a result, we can rewrite the Hamiltonian around $z_\alpha$ as

\begin{equation}
\begin{split}
    \tilde{H}_{susy}(g) = g \left[\sum_i\left(\frac{p_i^2}{2}+\frac{(\sum_j a_{i,j,\alpha}x_j)^2}{2}\right) \right.\\
    \left.
    + \frac{i}{2}\sum_{i,j}\gamma_{a,i}a_{j,i,\alpha}\gamma_{b,j} \right]\\
    + g^{1/2}\left[\sum_{i,j,k,l}\frac{a_{i,j,\alpha}b_{i,k,l,\alpha} x_j x_k x_l}{2}
    \right.
    \\    \left.
    + \frac{i}{2}\sum_{i,j,k}\gamma_{a,i}b_{j,i,k,\alpha}x_k\gamma_{b,j}\right] + \mathcal{O}(1)
\end{split}
\end{equation}
Therefore, the energy lifted or lowered due to the higher order corrections terms is of the order of $g^{1/2}$. As a result, if we choose large enough $g$, we can always guarantee that an energy window from $- \frac{1}{2}\min(\{g\lambda_{\alpha,i}\})$ to $\frac{1}{2}\min(\{g\lambda_{\alpha,i}\})$ only contains the $N$ states that have almost or exact zero-energy as shown in Fig.\ref{fig:sp2}. These $N$ states have exactly zero energy if we only consider the first two lines in $\tilde{H}_{susy}(g)$.

Then we can calculate the Witten index by only focusing on these $N$ states because all other nonzero energy states are paired, and thus contributes zero to the Witten index. From the result of linear-constraint cases, the fermion parity of the lowest energy state at $z_\alpha$ is $\sgn[\det(\frac{\partial f_i}{\partial x_j})]$. As a result, $W=\sum_{\alpha} \sgn[\det(\frac{\partial f_i}{\partial x_j}|_{z_\alpha})]$ which is exactly equal to $Q_{net}$.

In the next step, we are going to show that the Witten index $W$ is independent of $g$, namely, $\frac{dW}{dg}=0$. First, we write
\begin{equation}
\begin{split}
   W=\sum_{m_1}\langle  B_{m_1}|e^{-\beta H_{susy}(g)}|B_{m_1}\rangle  \\
   -\sum_{m_2}\langle  F_{m_2}|e^{-\beta H_{susy}(g)}|F_{m_2}\rangle 
\end{split}
\end{equation}
where $|B_{m_1}\rangle $ is an even fermion parity state and $|F_{m_2}\rangle $ is an odd fermion parity state. Then we can calculate
\begin{equation}
\begin{split}
   \frac{dW}{dg}=-\beta\sum_{m_1}\langle  B_{m_1}|\frac{dH_{susy}(g)}{dg}e^{-\beta H_{susy}(g)}|B_{m_1}\rangle  \\
   +\beta\sum_{m_2}\langle  F_{m_2}|\frac{dH_{susy}(g)}{dg}e^{-\beta H_{susy}(g)}|F_{m_2}\rangle 
\end{split}
\end{equation}
The change of the Witten index due to the change of states is zero because the Witten index is independent of basis. Mathematically, we can write 
\begin{equation}
\begin{split}
\frac{d\langle  m|}{dg}e^{-\beta H_{susy}(g)}|m\rangle +\langle  m|e^{-\beta H_{susy}(g)}\frac{d|m\rangle }{dg} \\
=e^{-\beta E_m}\frac{d\langle  m|m\rangle }{dg}=0
\end{split}
\end{equation}
Then use the fact that $\frac{dH_{susy}(g)}{dg}=2\{\frac{dQ}{dg},Q\}$ where $Q=\sum_{i}[(\psi_i(p_i+igf_i)+\psi_i^{\dagger}(p_i-igf_i)]$, we get
\begin{equation}
\begin{split}
   \frac{dW}{dg}=-2\beta\sum_{m_1}\langle  B_{m_1}|\left\{\frac{dQ}{dg},Q\right\}e^{-\beta H_{susy}(g)}|B_{m_1}\rangle  \\
   +2\beta\sum_{m_2}\langle  F_{m_2}|\left\{\frac{dQ}{dg},Q\right\}e^{-\beta H_{susy}(g)}|F_{m_2}\rangle 
\end{split}
\end{equation}
This expression only receives contribution from states $|B_{m_1}\rangle $ and $|F_{m_2}\rangle$ with finite energies, which always come in pairs.
A pair of states $|B_m\rangle $ and $|F_m\rangle $ are related by $\sqrt{\frac{E_m}{2}}|F_m\rangle =Q|B_m\rangle $ and $\sqrt{\frac{E_m}{2}}|B_m\rangle =Q|F_m\rangle $ where $E_m \neq 0$ is the eigenenergy of this pair of states. A unpaired state must be annihilated by $Q$ and, hence, has zero energy. Thus, we have
\begin{equation}
\begin{split}
   \frac{dW}{dg}=-\sqrt{2E_m}\beta\sum_{m}\langle  F_{m}|\frac{dQ(g)}{dg}e^{-\beta H_{susy}(g)}|B_{m}\rangle  \\
   -\sqrt{2E_m}\beta\sum_{m}\langle  B_{m}|\frac{dQ(g)}{dg}e^{-\beta H_{susy}(g)}|F_{m}\rangle  \\
   +\sqrt{2E_m}\beta\sum_{m}\langle  B_{m}|\frac{dQ(g)}{dg}e^{-\beta H_{susy}(g)}|F_{m}\rangle  \\
   +\sqrt{2E_m}\beta\sum_{m}\langle  F_{m}|\frac{dQ(g)}{dg}e^{-\beta H_{susy}(g)}|B_{m}\rangle 
\end{split}
\end{equation}
which is zero as long as $\frac{dQ(g)}{dg}$ is a regular function. In our case, $\frac{dQ(g)}{dg}=i\sum_i(\psi_i-\psi_i^{\dagger})f_i$, and thus $\frac{dW}{dg}=0$. Now, we have shown that for any constraint functions $\mathbf{f}$ that satisfy the three conditions, $Q_{net}=W$. 

Conceptually, we start with a purely classical system described by Eq. \ref{eq:Hiso} which has some zero-energy classical configurations characterized by the topological index $Q_{net}$.
Now, imaging ``turning on" quantum mechanics by treating Eq. \ref{eq:Hiso} as a quantum mechanical Hamiltonian. Generically, due to the Heisenberg uncertain principle, we do not expect any zero-energy eigenstates for Eq. \ref{eq:Hiso} anymore. However, we can ``recover" the zero-energy states by further including extra Majorana fermions interacting with the existing bosons and considering the Hamiltonian Eq. \ref{eq:Hsusy}.
% After turning on quantum mechanics, the energy of these zero-energy configuration states will be lifted due to the Heisenberg uncertain principle. Nevertheless, we can recover these zero-energy states by adding Majorana fermions that interacts with existing bosons. 
%In a hard constraint case such as stiff springs where $g \rightarrow \infty$, there are same number of zero-energy states corresponding to all zero-energy classical configurations.
%\cmj{[Comment: When we say ``stiff springs", are we talking about the classical model or the quantum model? So far I only see $g$ showing up in the quantum model. But later, stiff spring refers to mechanical linkage?]} 
%As we soften the constraints(reduce the stiffness of springs by making $g$ finite), some of zero-energy states will be lifted because of the overlap of wavefunctions due to quantum tunnelling effect. 
As a result, there will be a few zero-energy states that protected by supersymmetry. The number of these zero-energy states is characterized by the Witten index $W$. Physically, we can make an analogy between the two topological numbers $Q_{net}$ and $W$ in the following way:
\begin{equation}
\begin{split}
    \text{ minimum \#~of~zero-energy configurations } \\ 
    =\text{ minimum \#~of~zero-energy states}
\end{split}
\end{equation}
In the example of the KL chain, there is no supersymmetry-protected zero-energy state in the quantum system described Eq. \ref{eq:Hsusy} analogy to the KL chain because $W=Q_{net}=0$.
%\pwl{when springs are stiff (like mechanical linkages)}, there are many zero-energy many-body states including those $p$-wave superconductors. Nevertheless, all these zero-energy states will be lifted \pwl{as we reduce the stiffness of springs}. Therefore, there is no supersymmetry-protected zero-energy states in the quantum system analogy to the KL chain.
%\cmj{[Comment: There is no calculation on the KL chain in this paper. Where does all these statement come from?]}

\section{conclusion}

We show metamaterials can be used to study the topology of interacting quantum materials with the aid of supersymmetry. Specifically, we map a classical constrained problem to Bogoliubov quasiparticles of a superfluid/superconductor coupled to a boson such as a phonon. Hence, classical metamaterials can be used to study some aspects of the most challenging problems in quantum condensed matter physics. 

Necessarily, the connection between classical metamaterials and quantum materials requires fine tuning. The Debye temperatures in real materials range from $\mathcal{O}(10)$ to $\mathcal{O}(10^3)$ $K$ could match the order of the hopping strength of electrons in some materials. If the phonon band structure is similar to the electron band structure, and we fine-tune the anharmonicity of the phonon to match the coupling between Majorana fermions and phonon, it is possible to realize such a supersymmeric quantum system that shares the same topology of a classical mechanical systems. Perhaps a search through a database of all materials may find some that approximately meet these conditions. But even if not, the connection may still prove useful for the fine tuned problems may provide insight into the general behavior of interacting metals and superconductors. 

Potentially, there are many possible ways of defining topological indices following the prescription in Ref.\cite{PhysRevLett.127.076802}. Perhaps studying connections between these topological indices and existing topological numbers in quantum theory, as we have done in this manuscript, may yield further connections between metamaterials and quantum materials. If so, classical metamaterials may provide explanations of otherwise inexplicable behavior of some quantum materials.
%, such as the existence of intertwined orders\cite{fradkin2015colloquium}.
%points out an alternative way to understand the topology of quantum systems.

\bibliographystyle{unsrt}
\bibliography{bib}

\begin{thebibliography}{10}

\bibitem{PhysRevB.94.165101}
Michael~J. Lawler.
\newblock Supersymmetry protected topological phases of isostatic lattices and
  kagome antiferromagnets.
\newblock {\em Phys. Rev. B}, 94:165101, Oct 2016.

\bibitem{PhysRevResearch.1.032047}
Jan Attig, Krishanu Roychowdhury, Michael~J. Lawler, and Simon Trebst.
\newblock Topological mechanics from supersymmetry.
\newblock {\em Phys. Rev. Research}, 1:032047, Dec 2019.

\bibitem{PhysRevResearch.3.023213}
Robert~H. Jonsson, Lucas Hackl, and Krishanu Roychowdhury.
\newblock Entanglement dualities in supersymmetry.
\newblock {\em Phys. Rev. Research}, 3:023213, Jun 2021.

\bibitem{lawler2013emergent}
Michael~J Lawler.
\newblock Emergent gauge dynamics of highly frustrated magnets.
\newblock {\em New Journal of Physics}, 15(4):043043, 2013.

\bibitem{tm1}
C.~L. Kane and T.~C. Lubensky.
\newblock Topological boundary modes in isostatic lattices.
\newblock {\em Nature Physics}, 10:39, Dec 2013.

\bibitem{tm2}
Bryan Gin-ge Chen, Bin Liu, Arthur~A. Evans, Jayson Paulose, Itai Cohen,
  Vincenzo Vitelli, and C.~D. Santangelo.
\newblock Topological mechanics of origami and kirigami.
\newblock {\em Phys. Rev. Lett.}, 116:135501, Mar 2016.

\bibitem{tm3}
D.~Zeb Rocklin, Bryan Gin-ge Chen, Martin Falk, Vincenzo Vitelli, and T.~C.
  Lubensky.
\newblock Mechanical weyl modes in topological maxwell lattices.
\newblock {\em Phys. Rev. Lett.}, 116:135503, Apr 2016.

\bibitem{tm4}
Leyou Zhang and Xiaoming Mao.
\newblock Fracturing of topological maxwell lattices.
\newblock {\em New Journal of Physics}, 20(6):063034, jun 2018.

\bibitem{tm5}
Adrien Saremi and Zeb Rocklin.
\newblock Controlling the deformation of metamaterials: Corner modes via
  topology.
\newblock {\em Phys. Rev. B}, 98:180102, Nov 2018.

\bibitem{origami2}
Krishanu Roychowdhury, D.~Zeb Rocklin, and Michael~J. Lawler.
\newblock Topology and geometry of spin origami.
\newblock {\em Phys. Rev. Lett.}, 121:177201, Oct 2018.

\bibitem{roychowdhury2018classification}
Krishanu Roychowdhury and Michael~J Lawler.
\newblock Classification of magnetic frustration and metamaterials from
  topology.
\newblock {\em Physical Review B}, 98(9):094432, 2018.

\bibitem{PhysRevLett.127.076802}
Po-Wei Lo, Christian~D. Santangelo, Bryan Gin-ge Chen, Chao-Ming Jian, Krishanu
  Roychowdhury, and Michael~J. Lawler.
\newblock Topology in nonlinear mechanical systems.
\newblock {\em Phys. Rev. Lett.}, 127:076802, Aug 2021.

\bibitem{10.4310/jdg/1214437492}
Edward Witten.
\newblock {Supersymmetry and Morse theory}.
\newblock {\em Journal of Differential Geometry}, 17(4):661 -- 692, 1982.

\bibitem{tft}
D.~Birmingham, M.~Blau, M.~Rakowski, and G.~Thompson.
\newblock Topological field theory.
\newblock {\em Physics Reports}, 209:129--340, 1991.

\bibitem{brst1}
C.~Becchi, A.~Rouet, and R.~Stora.
\newblock The abelian higgs kibble model, unitarity of the s-operator.
\newblock {\em Physics Letters B}, 52(3):344 -- 346, 1974.

\bibitem{brst2}
I.~V. Tyutin.
\newblock Gauge invariance in field theory and statistical physics in operator
  formalism, 1975.

\bibitem{brst3}
C~Becchi, A~Rouet, and R~Stora.
\newblock Renormalization of gauge theories.
\newblock {\em Annals of Physics}, 98(2):287 -- 321, 1976.

\bibitem{Kitaev_2001}
A~Yu Kitaev.
\newblock Unpaired majorana fermions in quantum wires.
\newblock {\em Physics-Uspekhi}, 44(10S):131--136, oct 2001.

\bibitem{fp}
L.D. Faddeev and V.N. Popov.
\newblock Feynman diagrams for the yang-mills field.
\newblock {\em Physics Letters B}, 25(1):29 -- 30, 1967.

\end{thebibliography}

\appendix

\section{Derivation of symmetric cases}

We use an approach similar to the Faddeev-Popov gauge-fixing procedure\cite{fp}. First, we generalize $Q_{net}$ to a family of sets of constraints $\mathbf{f}(\mathbf{x})+\mathbf{w}=\mathbf{0}$ where $\mathbf{w}$ is some constant vector. The net topological index $Q_{net}(\mathbf{w})$ depending on $\mathbf{w}$ is written as
\begin{equation}
\begin{split}
    Q_{net}(\mathbf{w}) = \sum_{\mathbf{f}(p_{\mathbf{w}})+\mathbf{w}=\mathbf{0}}\mu(p_{\mathbf{w}})= \sum_{\mathbf{f}(p_{\mathbf{w}})+\mathbf{w}=\mathbf{0}} \sgn\left[\det\left(\frac{\partial f_i}{\partial x_j}\right)\right]\\
    =\sum_{\mathbf{f}(p_{\mathbf{w}})+\mathbf{w}
    =\mathbf{0}} \Big|\det\left(\frac{\partial f_i}{\partial x_j}\right)\Big|^{-1}\det\left(\frac{\partial f_i}{\partial x_j}\right)\\
    =\int d\mathbf{x}\prod_{i=1}^n \delta(f_i+w_i)\det\left(\frac{\partial f_i}{\partial x_j}\right)
\end{split}
\label{eq:Qs}
\end{equation}
In the first line, we assume that all solution points are non-degenerate. In the last line, we replace the sum of Jacobian by an integration over delta functions. $Q_{net}(\mathbf{w})$ can also be calculated by drawing a lager sphere that encloses all solution points and calculating the integration of a differential form in Eq.\ref{eq:windingnumber}. Thus, $Q_{net}(\mathbf{w})$ only depends on the asymptotic behavior of $\mathbf{f}(\mathbf{x})+\mathbf{w}$ on the boundaries of $\mathbf{x}$ ($||\mathbf{x}|| \to \infty$).

In the next step, we compute the average of the net topological indices over this family of sets of constraints by using the the weight $\prod_{i=1}^n e^{-\frac{w_i^2}{2}}$. Then the average of the net topological indices is
\begin{equation}
\begin{split}
    Q_{ave} = \int \prod_{i=1}^n\frac{dw_i} {\sqrt{2\pi}} e^{-\frac{w_i^2}{2}}\left[\int d\mathbf{x}\prod_{i=1}^
    n\delta(f_i+w_i)\det\left(\frac{\partial f_i}{\partial x_j}\right)\right]
\end{split}
\end{equation}
Under the condition that $||\mathbf{f}(\mathbf{x})|| \rightarrow \infty$ on the boundaries of $\mathbf{x}$, the asymptotic behavior of $\mathbf{f}(\mathbf{x})$ is unchanged under any finite local deformations (e.g., the deformation $\tilde{\mathbf{f}}=\mathbf{f}+\mathbf{w}$). Therefore, when $||\mathbf{f}(\mathbf{x})|| \rightarrow \infty$ as $||\mathbf{x}|| \to \infty$, $Q_{net}(\mathbf{w})$ is independent of $\mathbf{w}$ and $Q_{ave}$ is equal to the original $Q_{net}$ in Eq.\ref{eq:Qnet}

Then after integrating over $w_i$ and writing $\det(\frac{\partial f_i}{\partial x_j})$ as an integral over complex Grassmann numbers, $Q_{ave}$ can be rewritten as
\begin{equation}
\begin{split}
    Q_{ave} = \int \frac{\mathbf{dx}\mathbf{d\Psi} \mathbf{d\bar\Psi}}{(i\sqrt{2\pi})^{n}} \exp\left[-\sum_{i=1}^n \frac{1}{2}f_i^2\right.\\
    \left. -i\sum_{i=1}^n \sum_{j=1}^n (\bar \Psi_{i} \frac{\partial f_i}{\partial x_j} \Psi_{j}) \right]
    \label{eq:Qave}
\end{split}
\end{equation}
where $\bar \Psi_i$ and $\Psi_i$ are complex Grassmann numbers. We can see that $Q_{ave}$ plays a similar role as the partition function.

To promote the classical theory to a quantum theory, we consider another similar constrained problem by replacing $\mathbf{f}(\mathbf{x})$ by $\frac{d\mathbf{x}}{d\tau} + \mathbf{f}(\mathbf{x})$ where $\tau$ is the imaginary time. Following the same approach, the new topological index can be written as
\begin{equation}
\begin{split}
    W = \int \mathbf{Dx} \mathbf{D\Psi} \mathbf{D\bar \Psi}  \exp\left(-\oint d\tau\left[\sum_{i=1}^n \frac{1}{2} \left(\frac{dx_i}{d\tau} + f_i\right)^2 \right.\right.\\
    \left.\left. +i\sum_{i=1}^n \sum_{j=1}^n \bar \Psi_{i} \left(\delta_{i,j}\frac{d}{d\tau}+\frac{\partial f_i}{\partial x_j}\right) \Psi_{j}\right]\right),
\end{split}
\label{eq:susy}
\end{equation}
All constants are absorbed in $\mathbf{Dx} \mathbf{D\Psi} \mathbf{D\bar \Psi}$. Here we emphasize that $W$ is not a regular partition function because it requires periodic boundary conditions along the imaginary time circle for both bosons and fermions.

When $\partial f_i/\partial x_j$ is symmetric, namely when $f_i = \frac{\partial V}{\partial x_i}$, the path integral Eq. \ref{eq:susy} describes a supersymmetric quantum mechanics model with BRST symmetry \cite{tft}. In the following, we review some key aspects of this supersymmetric quantum mechanics model with BRST symmetry. The discussion below follows Ref. \onlinecite{tft}. We assume $f_i = \frac{\partial V}{\partial x_i}$ from now on.

It can be shown that only the configurations with $\frac{d\mathbf{x}}{d\tau} + \mathbf{f}(\mathbf{x})=\mathbf{0}$ contributes to the path integral. Naively, there can be
% With a set of constraints $\frac{d\mathbf{x}}{d\tau} + \mathbf{f}(\mathbf{x})=\mathbf{0}$, there are 
two types of solutions, dynamical solutions ($\frac{d\mathbf{x}}{d\tau}\neq 0$) and stationary solutions ($\frac{d\mathbf{x}}{d\tau}=0$). First, we notice that $\frac{d\mathbf{x}}{d\tau} + \mathbf{f}(\mathbf{x})=\mathbf{0}$ implies that
\begin{equation}
\begin{split}
    & 0=  \oint d\tau\sum_{i=1}^n \left(\frac{dx_i}{d\tau} + f_i\right)^2 \\
    & =\oint d\tau\sum_{i=1}^n \left(\frac{dx_i}{d\tau}\right)^2 + \oint d\tau\sum_{i=1}^n f_i^2 
    +2\oint d\tau\sum_{i=1}^n \frac{dx_i}{d\tau}f_i 
\end{split}
\end{equation}
Notice that $f_i=\frac{\partial V}{\partial x_i}$. the last term becomes $2\oint d\tau \frac{dV}{d\tau}$ which is zero due to the periodic boundary condition. Hence, $\frac{dx_i}{d\tau}=0$ and $f_i=0$, namely there are only stationary solutions.
For a stationary solution, the system stays at rest in a solution point $p$.
The fermion contribution to the topological index $W$ for each stationary solution
can be calculated by transforming the field to Fourier series. The sign only comes from the zero frequency term because nonzero frequency terms all comes in complex conjugate pairs and the product of a complex conjugate pair is always positive. Note the fermion has periodic boundary condition along the time direction, which permits zero-frequency modes.
Therefore, the total contribution from a stationary solution is the same as the topological index $\mu(p)$ defined in Eq. \ref{eq:windingnumber}. As a result, $W$ is equal to $Q_{ave}$ when $f_i=\frac{\partial V}{\partial x_i}$.

The BRST formulation can be recovered by adding auxiliary field $\mathbf{B}$. We rewrite $W$ as
\begin{equation}
\begin{split}
   & W = \int \mathbf{Dx} \mathbf{D\Psi} \mathbf{D\bar \Psi} \mathbf{DB}  \exp \left(-\oint \left[d\tau\sum_{i=1}^n \frac{1}{2}B_i^2 \right. \right. \\
    & \left. \left.
    -i\sum_{i=1}^n B_i\left(\frac{dx_i}{d\tau} + f_i\right) 
    +i\sum_{i=1}^n \sum_{j=1}^n \bar \Psi_{i} \left(\delta_{i,j}\frac{d}{d\tau}+\frac{\partial f_i}{\partial x_j}\right) \Psi_{j}\right]\right)
\end{split}
\end{equation}
The supersymmetry relation is defined via a nilpotent generator $Q=\sum_{i=1}^n \Psi_i B_i$. The transformation rules are
\begin{equation}
\{Q,x_i\}=\Psi_i \quad \{Q,B_i\}=0 \quad \{Q,\Psi_i\}=0 \quad \{Q,\bar \Psi_i\}=B_i
\label{eq:trans}
\end{equation}

% We can further write the topological index $W$ in the Hamiltonian formalism with a supersymmertic Hamiltonian
% \begin{equation}
% H_{BRST}=-\frac{1}{2} \sum_{i=1}^n \left(\frac{dx_i}{d\tau}\right)^2 + \frac{1}{2} \sum_{i=1}^n f_i^2 + i \sum_{i=1}^n \sum_{j=1}^n \bar \Psi_{i}\frac{\partial f_i}{\partial x_j}\Psi_{j}
% \end{equation}
% When $f_i=\frac{\partial V}{\partial x_i}$, in the Euclidean quantum theory, we can replace $i\frac{dx_i}{d\tau}$ by momentum operator $p_i$ and $\bar \Psi_{i}$ by $i\Psi_{i}^{\dagger}$. 
% \

The Hamiltonian of the BRST-symmetric model can then be written as
\begin{equation}
\begin{split}
H_{BRST} & =\frac{1}{2} \sum_{i=1}^n p_i^2 + \frac{1}{2} \sum_{i=1}^n f_i^2 \\
& + \frac{1}{2} \sum_{i=1}^n \sum_{j=1}^n \frac{\partial f_j}{\partial x_i}(\Psi_{i}\Psi_{j}^{\dagger}-\Psi_{i}^{\dagger}\Psi_{j})
\end{split}
\end{equation}
%\cmj{
%[Comment: Why is there an extra $-$ sign for the fermion part compared to (9)?. Maybe it is just a question about how the complex fermion operators is defined using the Majorana operators. But we should keep it consistent]} \cmj{Note that $\frac{\partial f_i}{\partial x_j} = \frac{\partial^2 V}{\partial x_i \partial x_j}$, which ensures the Hermiticity of the Hamiltonian.}
%For the calculation purpose, we replace $p_i$ by $i\frac{\partial}{\partial x_i}$ and $\Psi_{i}^{\dagger}$ by $\frac{\partial}{\partial \Psi_{i}}$, and get
%\begin{equation}
%\begin{split}
%H_{BRST}=-\frac{1}{2} \sum_{i=1}^n \frac{\partial^2}{\partial %x_i^2} + \frac{1}{2} \sum_{i=1}^n f_i^2 \\
%-\frac{1}{2} \sum_{i=1}^n \sum_{j=1}^n \frac{\partial f_i}{\partial %x_j}(\frac{\partial}{\partial %\Psi_{i}}\Psi_{j}+\Psi_{j}\frac{\partial}{\partial \Psi_{i}})
%\end{split}
%\label{eq:Hbrst}
%\end{equation}
In the Hamiltonian formalism, the topological index $W$ can be calculated by taking the trace or summing over eigenstates.
%\begin{equation}
%    W = Tr(e^{-\oint d\tau H_{BRST}})
%\end{equation}
%\cmj{[Comment: I think this notation is does not make sense. It looks like $Tr(e^{-\beta H_{BRST}})$, which would just the regular  function, not the Witten index. We can just simply say, with $H_{BRST}$, we can define the Witten index Eq. \ref{eq:Witten} ]}
%The total fermion number $n_f$ in $H_{BRST}$ is conserved, so we %can divide the eigenstates into different sections with a quantum %number $n_F$. 
For each fermion, there will an extra $\pi$ phase as a manifestation of the periodic boundary condition along the time circle in the path integral. As a result, the topological index $W$ is
\begin{equation}
    W = \sum_{m}(-1)^{n_F}e^{-\beta E_{m}}=\sum_{E_{m}=0}(-1)^{n_F}
\label{eq:Witten}
\end{equation}
which is indeed the Witten index. 
%\cmj{[Comment: Witten index is NOT the usual partition function. Eq. \ref{eq:Witten} and Eq. \ref{eq:susy} are different things. The partition function is defined as $\sum_{m} e^{-\beta E_{m}}$ and the Witten index is $\sum_{m}(-1)^{n_F}e^{-\beta E_{m}}$ ] }

\end{document}